# Direct Imaging of Nanoscale Conductance Evolution in Ion-Gel-Gated Oxide Transistors

Yuan Ren[1]*, Hongtao Yuan[2,3]*, Xiaoyu Wu[1], Zhuoyu Chen[2], Yoshihiro Iwasa[4], Yi Cui[2,3], Harold Y. Hwang[2,3], Keji Lai[1]

[1] *Department of Physics, University of Texas at Austin, Austin, Texas 78712, USA*

[2] *Geballe Laboratory for Advanced Materials, Stanford University, Stanford, California 94305, USA*

[3] *Stanford Institute for Materials and Energy Sciences, SLAC National Accelerator Laboratory, Menlo Park, California 94025, USA*

[4] *Quantum-Phase Electronics Center and Department of Applied Physics, University of Tokyo, Tokyo 113-8656, Japan and RIKEN Center for Emergent Matter Science, Wako 351-0198, Japan*

*These authors contributed equally to this work

**Abstract**

Electrostatic modification of functional materials by electrolytic gating has demonstrated a remarkably wide range of density modulation, a condition crucial for developing novel electronic phases in systems ranging from complex oxides to layered chalcogenides. Yet little is known microscopically when carriers are modulated in electrolyte-gated electric double-layer transistors (EDLTs) due to the technical challenge of imaging the buried electrolyte-semiconductor interface. Here, we demonstrate the real-space mapping of the channel conductance in ZnO EDLTs using a cryogenic microwave impedance microscope. A spin-coated ionic gel layer with typical thicknesses below 50 nm allows us to perform high resolution (on the order of 100 nm) sub-surface imaging, while maintaining the capability of inducing the metal-insulator transition under a gate bias. The microwave images vividly show the spatial evolution of channel conductance and its local fluctuations through the transition, as well as the uneven conductance distribution established by a large source-drain bias. The unique combination of ultra-thin ion-gel gating and microwave imaging offers a new opportunity to study the local transport and mesoscopic electronic properties in EDLTs.

Metal-insulator transitions (MITs), in which the electrical conductivity changes by orders of magnitude[1], are intriguing phenomena that underlie many long-standing physics problems such as unconventional superconductivity[2] and colossal magnetoresistance[3]. The same process is also important for modern semiconductor devices, in which the conductivity is controlled by either chemical doping or electrostatic field effect[4]. Using ionic liquids[5–7] or gels[8–11] as the gate dielectrics, the electric double-layer transistors (EDLTs) developed in the past few years have demonstrated the ability to modulate the sheet carrier density up to a level much higher than that achieved in conventional metal-oxide-semiconductor field-effect transistors (MOSFETs)[5–12]. Such unprecedented tunability of electron concentration represents a paradigm shift in condensed matter physics research since many carrier-mediated processes, previously only accessible through chemical substitution, can now be studied in the FET configuration with better controllability and less disorder effect[13, 14]. To date, the EDLT structure has been utilized to investigate the field-induced MITs[12, 15–17], magnetic ordering[18, 19], interfacial superconductivity[20–25], and topological surface states[19, 26] in a variety of advanced materials, with an ever increasing list in the foreseeable future.

The remarkable success in the EDLT research has mostly been made possible by transport[5–12, 15–26] and optical[27–29] experiments, in which macroscopic properties of the buried conduction channel at the EDL interfaces are measured. In contrast, it remains technically challenging to obtain the microscopic electrical information in the EDLT channel, which is crucial for the study of nanoscale electronic inhomogeneity in complex quantum materials[30–33] and the local conductance fluctuation in technologically important semiconductors. In conventional EDLT devices, however, the electrolyte-semiconductor interface is usually buried underneath a large droplet of ionic liquid/gel and thus cannot be directly studied by surface-sensitive electrical probes such as conductive atomic-force microscope (C-AFM) or scanning tunneling microscope (STM)[34]. In recent years, a number of local probes, including scanning Kelvin probe microscopy (SKPM)[35], electrostatic force microscopy (EFM)[36], scanning charge modulation microscopy (SCMM)[37], and Raman microscopy[38], have been utilized to image the FETs in a non-contact manner. These techniques, however, only provide indirect rather than direct information on the nanoscale conductance evolution. A new approach capable of spatially resolving the sub-surface electrical conductivity is therefore highly desirable to advance our knowledge on the electrostatic control of novel materials.

In this Letter, we report the first real-space electrical imaging of the channel conductance in an oxide EDLT by combining the cryogenic microwave impedance microscopy (MIM, see Supporting Information S1)[39–42] and electrolytic gating with ultra-thin ionic gels (thickness < 50 nm), as schematically illustrated in Fig. 1(a). The MIM is a powerful tool to detect the local permittivity and conductivity with a spatial resolution determined by the tip diameter (~100 nm) rather than the electromagnetic wavelength[43]. Thanks to the long-range tip-sample coupling[44], the MIM is capable of performing sub-surface imaging in the presence of a thin dielectric capping layer[41], which is ideal for measuring the ion-gel capped EDLT devices. We have observed the systematic evolution of local channel conductance during the insulator-to-metal transition induced by electrolytic gating. The uneven channel conductance in the presence of a large source-drain bias can also be imaged by MIM and the results are further corroborated by transport measurements and numerical simulations.

As a prototypical semiconducting oxide extensively studied in the EDLT configuration[6, 12, 45, 46], ZnO was used as the material platform in our experiment. The EDLT channel was defined by a pair of T-shaped Au electrodes for easy comparison between macroscopic transport and microscopic imaging, as shown in the scanning electron micrograph (SEM) in the inset of Fig. 1(b). The side metal gate was deposited on a thin layer of $Al_2O_3$ as the insulating spacer, which covered half of the ZnO surface. Details of the device fabrication and ionic gels are described in the Methods section. Fig. 1(b) shows the typical transfer characteristics of ion-gel-gated EDLTs with a source-drain bias $V_{DS}$ = 10 mV. The measurements were performed at 230 K, which is higher than the glass transition temperature ($T_g$ ~ 180 K) [12] to allow ionic motion but is low enough to avoid large leakage and irreversible electrochemical effects[46]. The gate dependence of the source-drain current ($I_{DS}$) clearly indicates that the EDLT can be turned on beyond a threshold voltage $V_{th}$ ~ 1.5 V with a negligible gate leakage current ($I_G$) below 2 nA. Note that we have studied samples with other contact configurations and the results were qualitatively the same (Supporting Information S2).

The good transfer characteristics and small leakage current suggest that the ZnO EDLT gated by an ultrathin ionic gel can still function as a normal transistor and induce the insulator-to-metal transition inside the channel. However, it should be addressed that the gate response of this device is relatively slow, as compared to ionic-liquid-gated EDLTs. While the actual distribution of ions in the gel can be quite complicated, an order-of-magnitude estimate of the gate response time can be obtained by examining the equivalent circuit in the inset of Fig. 1(c). At T = 230 K, the series resistance of the ultrathin ionic gel ($R_{IG}$) estimated from the gate current is about $10^9$ ~ $10^{10}$ Ω, much higher than that of the bulk ionic liquids/gels. Assuming a capacitance per unit area on the order of 10 µF/cm$^2$, as in previously reported ZnO EDLTs[12], the effective electric-double-layer capacitance $C_{EDL}$ of this millimeter-sized device is about 10$^{-6}$ F. As a result, the time constant ($\tau = R_{IG} \cdot C_{EDL} = 10^3 \sim 10^4$ sec) of this ion-gel-gated EDLT is very long, i.e., it typically took several hours or even longer for $I_{DS}$ to reach the true equilibrium state, which was indeed observed in the time dependence of $I_{DS}$ in Fig. 1(c).

Such a slow charging process owing to the small thickness of the gel layer has a direct consequence on our imaging experiment. For systems with a short response time, one usually measures the properties under equilibrium by varying the external $V_G$, which is difficult to realize here. Alternatively, taking advantages of the slow relaxation in our ion-gel-gated device, we may freeze the ionic motion at intermediate states by cooling the device below Tg of the gel, and study the corresponding microscopic distribution of local conductance in the EDLT channel. In our experiment, a $V_G$ of 2 V was applied at 230 K for various periods of waiting time until the desired $I_{DS}$ was reached. The sample was then cooled down for T-dependent measurements, as plotted in Fig. 2(a).

For each transport curve in Fig. 2(a), we acquired the MIM images near the source/drain electrodes at T = 100 K [Fig. 2(b)], with selected line scans shown in Supporting Information S3. For simplicity, only the imaginary (MIM-Im) components of the data, which fully capture the local conductance information, are displayed in Fig. 2(b). The real part (MIM-Re) of the MIM data can be found in Supporting Information S4. To obtain a quantitative understanding of the MIM images, we first present the finite-element analysis (FEA)[44] of the tip-sample interaction (Supporting Information S4), which converts the MIM-Im signals to the 2D sheet conductance $\sigma_{2D}$. Note that due to the generally non-negligible contact resistance and the specific source/drain geometry, one cannot directly calculate $\sigma_{2D}$ from the two-terminal conductance $G_{DS}$ measured by transport. In the following analysis, we only use $G_{DS}$ for an order-of-magnitude estimate when comparing with the $\sigma_{2D}$ maps. When the EDLT interface is insulating, the quasi-static microwave electric fields can spread into the bulk of ZnO. On the other hand, a highly conductive EDLT interface can effectively screen the microwave electric fields, which are then terminated at the ZnO surface. Consequently, the tip-sample capacitance, which is proportional to the MIM-Im signal, depends strongly on $\sigma_{2D}$. As plotted in Fig. 2(c), the MIM-Im signal remains low for small $\sigma_{2D}$, increases monotonically with increasing $\sigma_{2D}$ between 0.01 and 100 $\mu S \cdot sq$, and saturates for $\sigma_{2D}$ above 100 $\mu S \cdot sq$.

Of particular interest in Fig. 2(b) is the spatial evolution of local conductance as the EDLT channel was turned on. When a positive gate bias $V_G > V_{th}$ is applied, electrons in ZnO are first induced near source and drain by fringing fields from the cations accumulated on the electrodes. More cations are then attracted to these regions and the conductive area in ZnO gradually propagates into the entire channel. Such a process is vividly manifested by the MIM data. As $G_{DS}$ at 100 K increased from 0 to 1.5 $\mu S$, the light blue regions in the false-colored map ($\sigma_{2D} \sim 1$ $\mu S \cdot sq$) with higher MIM signals than the initial insulating state (dark blue, $\sigma_{2D} < 0.1$ $\mu S \cdot sq$) appeared around the source/drain contacts and propagated toward the center of the channel. For $G_{DS} = 3.4$ $\mu S$, the highly conductive areas (orange to red, $\sigma_{2D} \sim 10$ $\mu S \cdot sq$) originated from the two electrodes started to merge. At the highest $G_{DS}$ of 10 $\mu S$ in our experiment, the ZnO became highly conductive with $\sigma_{2D}$ > 10 µS · sq everywhere inside the channel. We emphasize that, while the same process is ubiquitous in FETs, it is a rare occasion that the evolution can be imaged by scanning probe experiments, which provide not only strong support on the effectiveness of the ion-gel gating but also the real-space information of field-induced MITs.

Evidence of the spatially inhomogeneous channel conductance has also been observed in our MIM experiment. As discussed before, the MIM-Im signals saturate for $\sigma_{2D}$ < $10^{-8}$ S · sq (insulating limit) and $\sigma_{2D}$ > $10^{-4}$ S · sq (conducting limit). Therefore, small fluctuations of local conductance are best visualized in the crossover regime [Fig. 3(a)] around $\sigma_{2D}$ = $10^{-6}$ S · sq. Fig. 3(b) shows the MIM images of a small area inside the channel at three different $G_{DS}$ values. For better visualization of the data, we remove the linear background that contains the absolute $\sigma_{2D}$ information and display the relative variations of MIM signals with a false-color scale different from that used in Fig. 2. As shown in Fig. 3(b), only several surface particles with low MIM signals (Supporting Information S4) were seen when the ZnO channel was insulating ($G_{DS}$ = 0 µS). For an intermediate $G_{DS}$ = 1.3 µS, which likely corresponds to an average $\sigma_{2D}$ around 1 µS · sq in the channel, appreciable mesoscopic conductance fluctuation could be observed in the image. A line profile in Fig. 3(c) shows that a spatial resolution of ~ 170 nm can be obtained in this regime. Due to the saturation of MIM response at the conducting limit, the image acquired at $G_{DS}$ = 6.7 µS again shows spatially uniform signals except for the same surface particles described before. Further experiments are needed to elucidate the origin and evolution of these non-uniform states. Nevertheless, the ability of resolving electrical inhomogeneity during the MITs will be particularly useful for the study of strongly correlated systems with nanoscale phase separation[30].

In order to further demonstrate the MIM imaging on EDLTs, we studied the local conductance profile under two different VDS's applied across the channel. For a small VDS of 10 mV in Fig. 4(a), both the ions and induced electrons were evenly distributed in the channel, which agreed well with the MIM conductance map in Fig. 4(b). In contrast, when a large bias $V_{DS}$ = 2 V was applied at 230 K, an asymmetric lateral conductance distribution was set up along the channel [Fig. 4(d)], reminiscent of the textbook description of FETs in the saturation regime[4]. The corresponding MIM image in Fig. 4(e) clearly shows that the conductive regions were pushed away from the drain electrode. Interestingly, since ions are immobilized upon cooling through the glass transition, the conductance landscape in the channel established by $V_G$ and $V_{DS}$ above $T_g$ will be frozen in space below $T_g$. This effect could have a strong impact on the transport characteristics, as recently reported in EDLTs fabricated on $MoS_2$ and $WSe_2$ flakes[29, 47, 48]. For the bias configuration in Fig. 4(a), appreciable $I_{DS}$ in the µA range was measured [Fig. 4(c)] when $V_{DS}$ was swept between –1 V and 1 V at 100 K, presumably due to the high $\sigma_{2D}$ in between the source/drain contacts. On the other hand, when the channel was asymmetrically biased at 230 K [Fig. 4(d)], a strong rectifying behavior was seen in the I-V characteristics in Fig. 4(f), again consistent with the low $\sigma_{2D}$ near the drain contact.

The uneven distribution of $\sigma_{2D}$ in the presence of a large VDS can be understood by considering the local surface band bending (SBB) inside the channel. In Figs. 5(a) and 5(b), we show the simulated depth profiles of the conduction band edge and total 3D electron density ($n_{3D}$) near the gel-ZnO interface with the SBB values of –0.4 eV, –1.0 eV, and –2.0 eV. Details of the Poisson-Schrodinger simulation[49] are provided in Supporting Information S5. The effective 2D density ($n_{2D}$) as a function of various SBB values, as plotted in Fig. 5(c), can then be calculated from the 3D density profile. Assuming a much larger gate area than the channel region and a low-temperature ZnO mobility of 100 cm$^2$/V·s, [12] we have self-consistently simulated the conductance distribution around the source/drain electrodes using a simple resistor-network model (Supporting Information S5). The resultant $\sigma_{2D}$ map, as shown in Fig. 5(d), agrees qualitatively with the MIM image in Fig. 4(e). Aided by the numerical analysis described above, our MIM and transport data can be built on solid ground for future investigations of electronic phase transitions in novel material systems.

In summary, we have, for the first time, demonstrated the electrical imaging of local channel conductance in ion-gel-gated oxide EDLTs by cryogenic microwave impedance microscopy. We found that electrons induced by the electrostatic field effect propagate from the source and drain electrodes to the center of the channel. Small fluctuations of the local conductance were also observed during the insulator-metal transition. By applying a large source-drain bias above the glass transition temperature of the gel, an uneven conductance profile was established across the EDLT channel, which was visualized by the MIM and further investigated by transport measurements and numerical simulations. The combination of ultra-thin ion-gel gating and microwave microscopy paves the way for studying the microscopic evolution of phase transitions in complex materials induced by electrostatic field effects.

**Methods**

**Device Fabrication.** The T-shaped source/drain electrodes were patterned on the bulk ZnO substrates (c-cut crystals from MTI Corporation) using standard electron beam lithography. A large area gold pad (60 nm thick), which served as the side gate electrode, was deposited on an $Al_2O_3$ (100 nm thick) isolation layer. Different from the vertical metal-dielectric-semiconductor structure in conventional MOSFETs, the configuration of side metal gating provides the opportunity for scanning probe microscopy directly from the top of the channel without being shadowed by a top metal electrode. An optimized ionic gel solution [DEME-TFSI-based, N,N-diethyl-N-(2-methoxyethyl)-N-methylammonium bis-trifluoromethylsulfonyl)-imide from Kanto Chemical Co.] was spin-coated on the device with a speed of 6000 rpm, followed by a vacuum baking at 80 ºC for 12 hours. As confirmed by our AFM measurement, the gel thickness was in the range of 30~50 nm, which is ideal for MIM imaging with a spatial resolution on the order of 100 nm.

**Microwave Impedance Microscopy and Transport Measurements.** The MIM setup for this experiment is based on a Janis ST-500 liquid helium flow cryostat, whose temperature can be varied continuously from room temperature down to about 20 K. Gate and source-drain bias voltages were applied through DC wires to the device. The transport data were acquired by two Keithley 2400 source meters. The micro-fabricated shielded probes[42] are commercially available from PrimeNano Inc. Finite-element modeling was performed using the commercial software COMSOL 4.4.

# REFERENCE

1. Imada, M., Fujimori, A. & Tokura, Y. Metal-insulator transitions. *Rev. Mod. Phys.* **1998**, 70, 1039 – 1263.

2. Lee, P. A., Nagaosa, N. & Wen, X.-G. Doping a Mott insulator: Physics of high-temperature superconductivity. *Rev. Mod. Phys.* **2006**, 78, 17 – 85.

3. Salamon, M. B. & Jaime, M. The physics of manganites: Structure and transport. *Rev. Mod. Phys.* **2001**, 73, 583 – 628.

4. Sze, S. M. & Ng, K. K. *Physics of Semiconductor Devices*. John Wiley& Sons, 3$^{rd}$ Edition, **2007**.

5. Misra, R., McCarthy, M. & Hebard, A. F. Electric field gating with ionic liquids. *Appl. Phys. Lett.* **2007**, 90, 052905.

6. Shimotani, H., Asanuma, H., Tsukazaki, A., Ohtomo, A., Kawasaki, M. & Iwasa Y. Insulator-to-metal transition in ZnO by electric double layer gating. *Appl. Phys. Lett.* **2007**, 91, 082106.

7. Simon, P. & Gogotsi, Y. Materials for electrochemical capacitors. *Nat. Mater.* **2008**, 7, 845 – 854.

8. Cho, J. H., Lee, J., He, Y., Kim, B., Lodge, T. P. & Frisbie C. D. High-Capacitance Ion Gel Gate Dielectrics with Faster Polarization Response Times for Organic Thin Film Transistors. *Adv. Mater.* **2008**, 20, 686 – 690.

9. Cho, J. H., Lee, J., Xia, Y., Kim, B., He, Y., Renn, M. J., Lodge, T. P. & Frisbie C. D. Printable ion-gel gate dielectrics for low-voltage polymer thin-film transistors on plastic. *Nat. Mater.* **2008**, 7, 900 – 906.

10. Yomogida, Y., Pu, J., Shimotani, H., Ono, S., Hotta, S., Iwasa, Y. & Takenobu, T. Ambipolar Organic Single-Crystal Transistors Based on Ion Gels. *Adv. Mater.* **2012**, 24, 4392 – 4397.

11. Lee, K. H., Kang, M. S., Zhang, S., Gu, Y., Lodge, T. P. & Frisbie C. D. "Cut and Stick" Rubbery Ion Gels as High Capacitance Gate Dielectrics. *Adv. Mater.* **2012**, 24, 4457 – 4462.

12. Yuan, H., Shimotani, H., Tsukazaki, A., Ohtomo, A., Kawasaki, M. & Iwasa, Y. High-Density Carrier Accumulation in ZnO Field-Effect Transistors Gated by Electric Double Layers of Ionic Liquids. *Adv. Funct. Mater.* **2009**, 19, 1046 – 1053.

13. Ahn, C. H., Triscone, J.-M. & Mannhart, J. Electric field effect in correlated oxide systems. *Nature* **2003**, 424, 1015 – 1018.

14. Ahn, C. H., Bhattacharya, A., Di Ventra, M., Eckstein, J. N., Frisbie, C. D., Gershenson, M. E., Goldman, A. M., Inoue, I. H., Mannhart, J., Millis, A. J., Morpurgo, A. F., Natelson, D. & Triscone, J.-M. Electrostatic modification of novel materials. *Rev. Mod. Phys.* **2006**, 78, 1185 – 1212.

15. Dhoot, A. S., Israe, C., Moya, X., Mathur, N. D. & Friend, R. H. Large Electric Field Effect in Electrolyte-Gated Manganites. *Phys. Rev. Lett.* **2009**, 102, 136402.

16. Nakano, M., Shibuya, K., Okuyama, D., Hatano, T., Ono, S., Kawasaki, M., Iwasa, Y. & Tokura, Y. Collective bulk carrier delocalization driven by electrostatic surface charge accumulation. *Nature* **2012**, 487, 459 – 462.

17. Jeong, J., Aetukuri, N., Graf, T., Schladt, T. D., Samant, M. G. & Parkin, S. S. P. Suppression of Metal-Insulator Transition in $VO_2$ by Electric Field–Induced Oxygen Vacancy Formation. *Science* **2013**, 339, 1402 – 1405.

18. Yamada, Y., Ueno, K., Fukumura, T., Yuan, H. T., Shimotani, H., Iwasa, Y., Gu, L., Tsukimoto, S., Ikuhara, Y. & Kawasaki, M. Electrically Induced Ferromagnetism at Room Temperature in Cobalt-Doped Titanium Dioxide. *Science* **2011**, 332, 1065 – 1067.

19. Checkelsky, J. G., Ye, J., Onose, Y., Iwasa, Y. & Tokura, Y. Dirac-fermion-mediated ferromagnetism in a topological insulator. *Nat. Phys.* **2012**, 8, 729 – 733.

20. Ueno, K., Shimotani, H., Yuan, H., Ye, J., Kawasaki, M. & Iwasa, Y. Field-Induced Superconductivity in Electric Double Layer Transistors. *J. Phys. Soc. Jpn.* **2014**, 83, 032001.

21. Ueno, K., Nakamura, S., Shimotani, H., Ohtomo, A., Kimura, N., Nojima, T., Aoki, H., Iwasa, Y. & Kawasaki M. Electric-field-induced superconductivity in an insulator. *Nat. Mater.* **2008**, 7, 855 – 858.

22. Ye, J. T., Inoue, S., Kobayashi, K., Kasahara, Y., Yuan, H. T., Shimotani H. & Iwasa Y. Liquid-gated interface superconductivity on an atomically flat film. *Nat. Mater.* **2010**, 9, 125 – 128.

23. Ueno, K., Nakamura, S., Shimotani, H., Yuan, H. T., Kimura, N., Nojima, T., Aoki, H., Iwasa, Y. & Kawasaki M. Discovery of superconductivity in $KTaO_3$ by electrostatic carrier doping. *Nat. Nanotech.* **2011**, 6, 408 – 412.

24. Bollinger, A. T., Dubuis, G., Yoon, J., Pavuna, D., Misewich J. & Bozovic I. Superconductor–insulator transition in $La_{2-x}Sr_xCuO_4$ at the pair quantum resistance. *Nature* **2011**, 472, 458 – 460.

25. Lee, Y., Clement, C., Hellerstedt, J., Kinney, J., Kinnischtzke, L., Leng, X., Snyder, S. D. & Goldman, A. M. Phase Diagram of Electrostatically Doped $SrTiO_3$. *Phys. Rev. Lett.* **2011**, 106, 136809.

26. Kim, D., Cho, S., Butch, N. P., Syers, P., Kirshenbaum, K., Adam, S., Paglione, J. & Fuhrer, M. S. Surface conduction of topological Dirac electrons in bulk insulating $Bi_2Se_3$. *Nat. Phys.* **2012**, 8, 459 – 463.

27. Yin, C., Yuan, H., Wang, X., Liu, S., Zhang, S., Tang, N., Xu, F., Chen, Z., Shimotani, H., Iwasa, Y., Chen, Y., Ge, W. & Shen, B. Tunable Surface Electron Spin Splitting with Electric Double-Layer Transistors Based on InN. *Nano Lett.* **2013**, 13, 2024 – 2029.

28. Yuan, H., Wang, X., Lian, B., Zhang, H., Fang, X., Shen, B., Xu, G., Xu, Y., Zhang, S.-C., Hwang, H. Y. & Cui, Y. Generation and electric control of spin–valley-coupled circular photogalvanic current in $WSe_2$. *Nat. Nanotech.* **2014**, 9, 851 – 857.

29. Zhang, Y. J., Oka, T., Suzuki, R., Ye, J. T. & Iwasa, Y. Electrically Switchable Chiral Light-Emitting Transistor. *Science* **2014**, 344, 725 – 728.

30. Dagotto, E. *Nanoscale phase separation and colossal magnetoresistance: the physics of manganites and related compounds*, Springer, New York, **2002**.

31. Martin, J., Akerman, N., Ulbricht, G., Lohmann, T., Smet, J. H., von Klitzing, K. & Yacoby, A. Observation of electron–hole puddles in graphene using a scanning single-electron transistor. *Nat. Phys.* **2008**, 4, 144 – 148.

32. Beidenkopf, H., Roushan, P., Seo, J., Gorman, L., Drozdov, I., Hor, Y. S., Cava R. J. & Yazdani, A. Spatial fluctuations of helical Dirac fermions on the surface of topological insulators. *Nat. Phys.* **2011**, 7, 939 – 943.

33. Mann, C., West, D., Miotkowski, I., Chen, Y. P., Zhang, S. & Shih, C.-K. Mapping the 3D surface potential in $Bi_2Se_3$. *Nat. Comm.* **2013**, 4, 2277.

34. Bonnell, D. A., Basov, D. N., Bode, M., Diebold, U., Kalinin, S. V., Madhavan, V., Novotny, L., Salmeron, M., Schwarz, U. D. & Weiss, P. S. Imaging physical phenomena with local probes: From electrons to photons. *Rev. Mod. Phys.* **2012**, 84, 1343 – 1381.

35. Burgi, L., Sirringhaus, H. & Friend, R. H. Noncontact potentiometry of polymer field-effect transistors. *Appl. Phys. Lett*. **2002**, 80, 2913 – 2915.

36. Muller, E. M. & Marohn, J. A. Microscopic Evidence for Spatially Inhomogeneous Charge Trapping in Pentacene. *Adv. Mater*. **2005**, 17, 1410 – 1414.

37. Sciascia, C., Martino, N., Schuettfort, T., Watts, B., Grancini, G., Antognazza, M. R., Zavelani-Rossi, M., McNeill, C. R. & Caironi, M. Sub-Micrometer Charge Modulation Microscopy of a High Mobility Polymeric n-Channel Field-Effect Transistor. *Adv. Mater*. **2011**, 23, 5086 – 5090.

38. Zaumseil, J., Jakubka, F., Wang, M. & Gannott, F. In Situ Raman Mapping of Charge Carrier Distribution in Electrolyte-Gated Carbon Nanotube Network Field-Effect Transistors. *J. Phys. Chem. C* **2013**, 117, 26361 – 26370.

39. Kundhikanjana, W., Lai, K., Kelly, M. A. & Shen, Z.-X. Cryogenic microwave imaging of metal–insulator transition in doped silicon. *Rev. Sci. Instrum.* **2011**, 82, 033705.

40. Lai, K., Nakamura, M., Kundhikanjana, W., Kawasaki, M., Tokura, Y., Kelly, M. A. & Shen, Z.-X. Mesoscopic Percolating Resistance Network in a Strained Manganite Thin Film. *Science* **2010**, 329, 190 – 193.

41. Lai, K., Kundhikanjana, W., Kelly, M. A., Shen, Z.-X., Shabani, J. & Shayegan, M. Imaging of Coulomb-Driven Quantum Hall Edge States. *Phys. Rev. Lett.* **2011**, 107, 176809.

42. Yang, Y., Lai, K., Tang, Q., Kundhikanjana, W., Kelly, M. A., Zhang, K., Shen, Z.-X. & Li, X. Batch-fabricated cantilever probes with electrical shielding for nanoscale dielectric and conductivity imaging. *J. Micromech. Microeng.* **2012**, 22, 115040.

43. Lai, K., Kundhikanjana, W., Kelly, M. A. & Shen, Z.-X. Nanoscale microwave microscopy using shielded cantilever probes. *Appl. Nanosci.* **2011**, 1, 13 – 18.

44. Lai, K., Kundhikanjana, W., Kelly, M. A. & Shen, Z.-X. Modeling and characterization of a cantilever-based near-field scanning microwave impedance microscope. *Rev. Sci. Instrum.* **2008**, 79, 063703.

45. Yuan, H., Shimotani, H., Tsukazaki, A., Ohtomo, A., Kawasaki, M. & Iwasa, Y. Hydrogenation-Induced Surface Polarity Recognition and Proton Memory Behavior at Protic-Ionic-Liquid/Oxide Electric-Double-Layer Interfaces. *J. Am. Chem. Soc.* **2010**, 132, 6672 – 6678.

46. Yuan, H., Shimotani, H., Ye, J., Yoon, S., Aliah, H., Tsukazaki, A., Kawasaki, M. & Iwasa, Y. Electrostatic and Electrochemical Nature of Liquid-Gated Electric-Double-Layer Transistors Based on Oxide Semiconductors. *J. Am. Chem. Soc.* **2010**, 132, 18402 – 18407.

47. Zhang, Y. J., Ye, J. T., Yomogida, Y., Takenobu, T. & Iwasa, Y. Formation of a Stable p–n Junction in a Liquid-Gated MoS$_2$ Ambipolar Transistor. *Nano Lett.* **2013**, 13, 3023 – 3028.

48. Yuan, H., Bahramy, M. S., Morimoto, K., Wu, S., Nomura, K., Yang, B.-J., Shimotani, H., Suzuki, R., Toh, M., Kloc, C., Xu, X., Arita, R., Nagaosa N. & Iwasa, Y. Zeeman-type spin splitting controlled by an electric field. *Nat. Phys.* **2013**, 9, 563 – 569.

49. Tan, I.-H., Snider, G. L., Chang, L. D. & Hu, E. L. A self-consistent solution of Schrödinger-Poisson equations using a nonuniform mesh. *J. Appl. Phys.* **1990**, 68, 4071.

# ASSOCIATED CONTENT

## Supplementary Information

It contains a more detailed description of the cryogenic microwave impedance microscopy, transport and MIM data on a different ion-gel-gated EDLT sample, full analysis of the MIM images, and Poisson-Schrodinger calculation at different surface band bending values. This material is available free of charge via the Internet at http://pubs.acs.org.

# AUTHOR INFORMATION

## Corresponding Author

Email: kejilai@physics.utexas.edu

## Notes

The authors declare no competing financial interests.

# ACKNOWLEDGEMENTS

We thank Chih-Kang Shih for valuable discussions. The MIM work at UT-Austin (Y.R., X.W. and K.L.) was supported by the US Department of Energy, Office of Science, Basic Energy Sciences under Early Career Award DE-SC0010308. The work at Stanford (H-T.Y., Z.C., Y.C., and H.Y.H.) was supported by the Department of Energy, Office of Basic Energy Sciences, Division of Materials Sciences and Engineering, under contract DE-AC02-76SF00515. Y. I. was supported by JSPS Grant-in-Aid for Specially Promoted Research (No. 25000003).

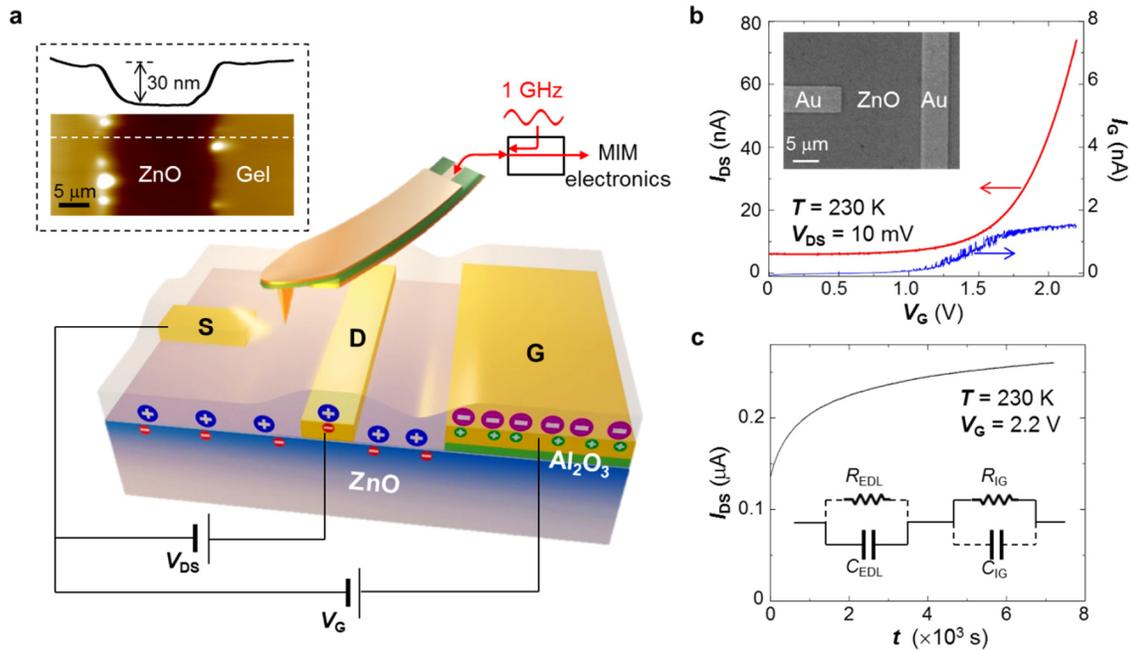

**Figure 1.** (**a**) Schematic diagram of the EDLT device and the MIM setup. The 1 GHz microwave signal is guided to the shielded cantilever probe and the reflected signal is detected by the MIM electronics. The inset shows the AFM image and a line cut of a typical spin-on ionic gel layer (thickness ∼ 30 nm), whose middle part was scratched away for thickness measurement. (**b**) Transfer characteristics of the EDLT measured at 230 K with $V_{DS}$ = 10 mV. The inset shows the SEM image of the ZnO channel defined by a pair of T-shaped Au contacts. (**c**) Slow relaxation of the source-drain current at 230 K. The equivalent circuit of the charging process is shown in the inset. For the ultra-thin ionic gel, the dominant circuit elements are the capacitance of the electric double layer $C_{EDL}$ and the resistance of the ionic gel $R_{IG}$, resulting in a long time constant on the order of $10^4$ sec.

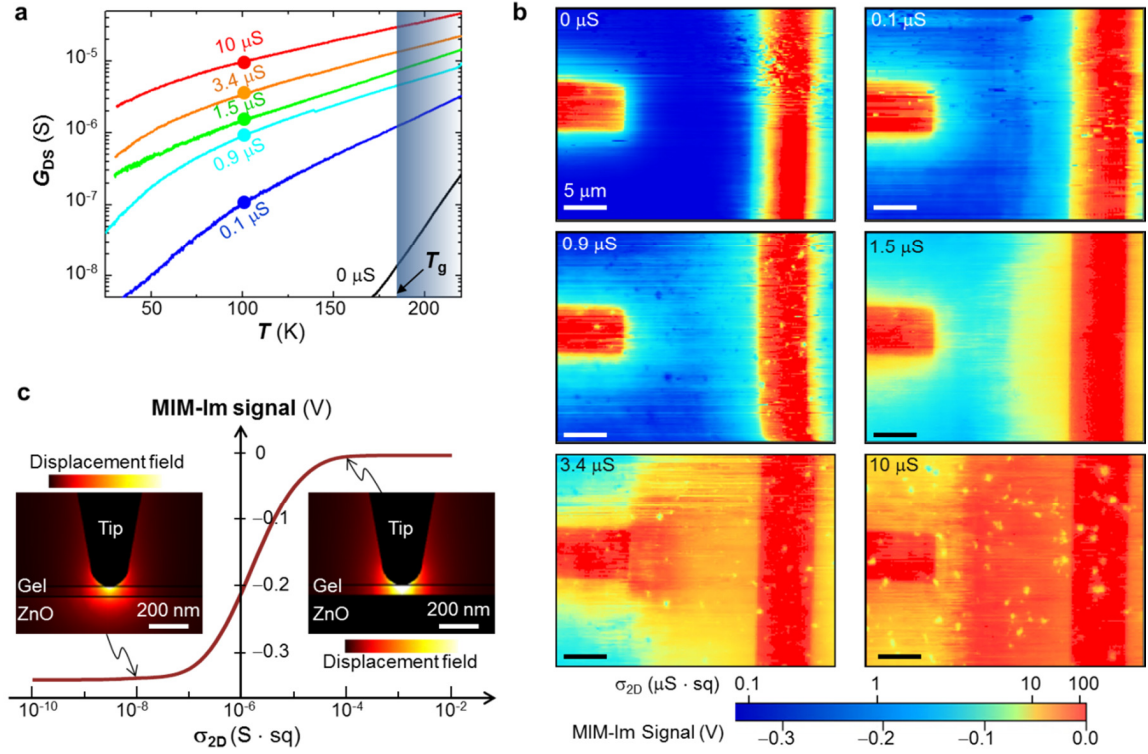

**Figure 2.** (**a**) Temperature dependence of the channel conductance $G_{DS}$ ($V_{DS}$ = 10 mV) as the device was gradually turned on. The $G_{DS}$'s at 100 K are labeled on each curve. As temperature decreased from 230 K (shaded region), the ionic motion slowed down and completely stops below $T_g$ ~ 180 K. Electron transport through the ZnO surface, however, was not affected by the glass transition of the ionic gel. (**b**) MIM-Im images with different $G_{DS}$'s at 100 K. Some surface particles with lower MIM signals, whose locations change from cool-down to cool-down, can be seen on the device (Supporting Information S3). The false color scale shows both the measured MIM signals and the 2D sheet conductance $\sigma_{2D}$ simulated from finite-element analysis (FEA). All scale bars are 5 μm. (**c**) FEA simulation of MIM-Im signals as a function of $\sigma_{2D}$ at the gel-ZnO interface. The maps of the quasi-static 1 GHz displacement field amplitude (D = εE, where ε is the permittivity and E the electric field) at the insulating (left, $\sigma_{2D} < 10^{-8}$ S · sq) and conducting (right, $\sigma_{2D} > 10^{-4}$ S · sq) limits are also shown in the insets. Scale bars in the insets are 200 nm.

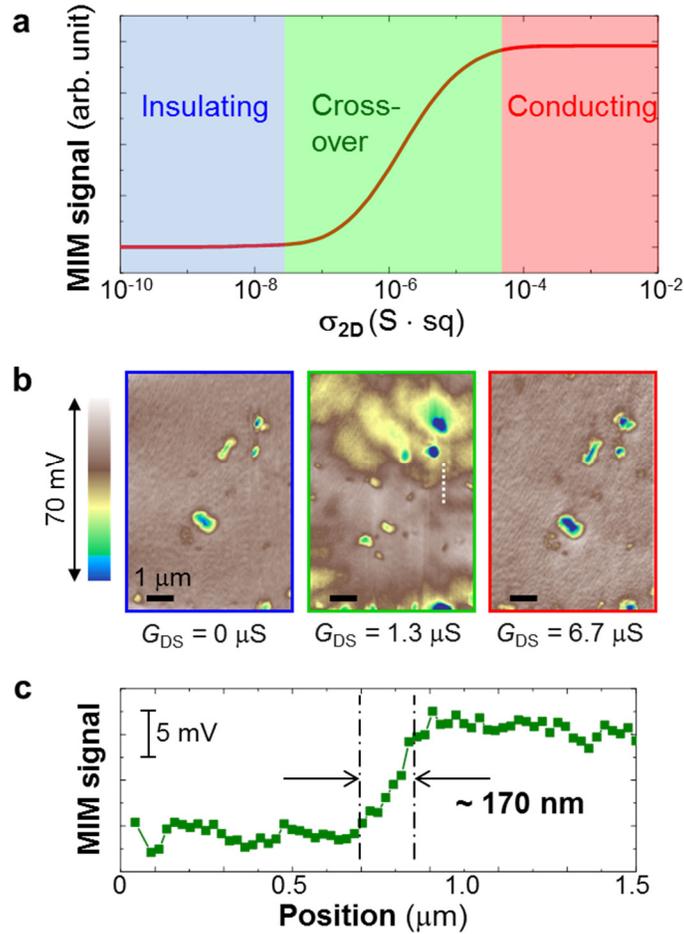

**Figure 3.** (**a**) Simulated results of MIM-Im signals as a function of $\sigma_{2D}$. The insulating, crossover, and conducting regimes are color-coded as blue, green, and red, respectively. (**b**) MIM images at three different $G_{DS}$'s after removing the background signals. Only several surface particles fixed in location were seen in the data when the ZnO channel was insulating ($G_{DS}$ = 0 μS) or relatively conducting ($G_{DS}$ = 6.7 μS). Fluctuation of the local conductance was observed when $G_{DS}$ = 1.3 μS, which is likely within the crossover regime. All scale bars are 1 μm. (**c**) A line profile (white dotted line in **b**) in the crossover regime, showing a spatial resolution of about 170 nm for the MIM signals.

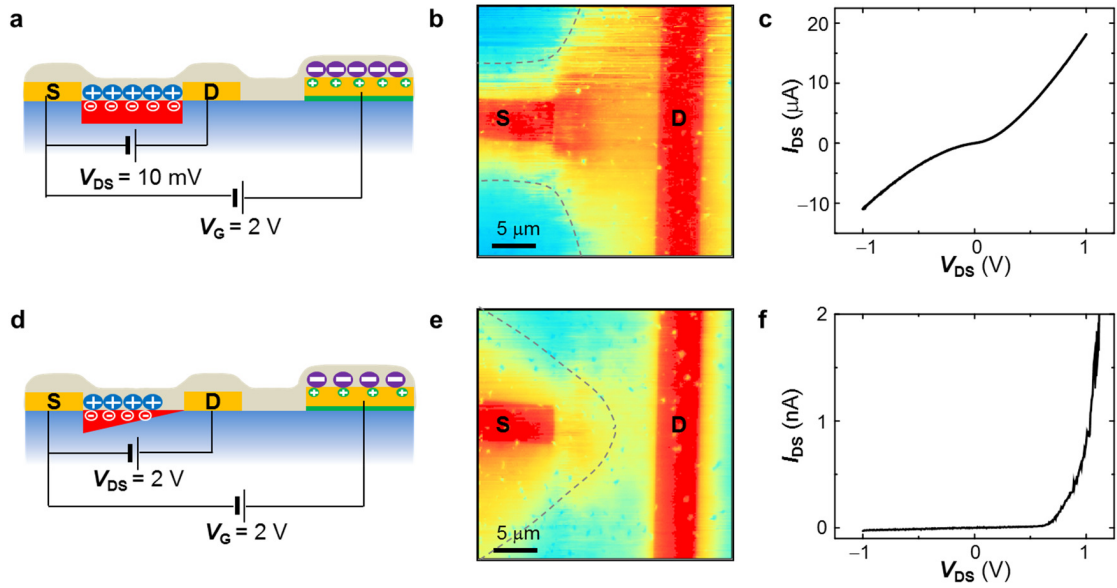

**Figure 4.** (**a**) Schematic of the EDLT and the charge distribution with a small $V_{DS}$ = 10 mV. (**b**) Corresponding MIM image at 100 K around the source and drain electrodes. (**c**) $I_{DS}$-$V_{DS}$ characteristics when the device under the bias condition in (**a**) was cooled to 100 K. (**d – f**) Same as (**a – c**) except that a large $V_{DS}$ = 2 V was applied at 230 K before cooling down to 100 K for the MIM imaging and transport measurement. The dashed lines are guides to the eyes for the boundaries of conductive regions (yellow to red in the false color map). The scale bars are 5 μm.

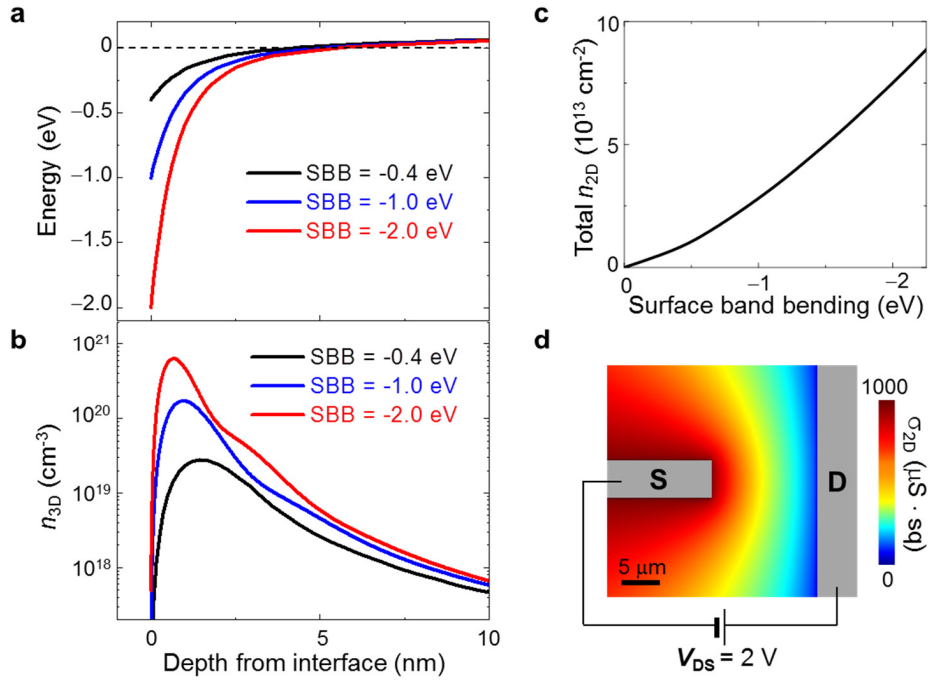

**Figure 5.** (**a**) Poisson-Schrodinger simulation result of the conduction band edge and (**b**) 3D electron density ($n_{3D}$) as a function of the depth from the EDLT interface. The results under three surface band bending (SBB) values of –0.4 eV, –1.0 eV, and –2.0 eV are plotted. (**c**) Total 2D density ($n_{2D}$) as a function of the SBB. (d) Simulated local conductance distribution induced by a large $V_{DS}$ = 2 V across the EDLT channel.

# Supporting Information

## Direct Imaging of Nanoscale Conductance Evolution in Ion-Gel-Gated Oxide Transistors

Yuan Ren[1*], Hongtao Yuan[2,3*], Xiaoyu Wu[1], Zhuoyu Chen[2,3], Yoshihiro Iwasa[4], Yi Cui[2,3], Harold Y. Hwang[2,3], Keji Lai[1]

1. Department of Physics, University of Texas at Austin, Austin, Texas 78712, USA

2. Geballe Laboratory for Advanced Materials, Stanford University, Stanford, California 94305, USA

3. Stanford Institute for Materials and Energy Sciences, SLAC National Accelerator Laboratory, Menlo Park, California 94025, USA

4. Quantum-Phase Electronics Center and Department of Applied Physics, University of Tokyo, Tokyo 113-8656, Japan and RIKEN Center for Emergent Matter Science, Wako 351-0198, Japan

* These authors contributed equally to this work

**Section 1: Cryogenic microwave impedance microscopy.**

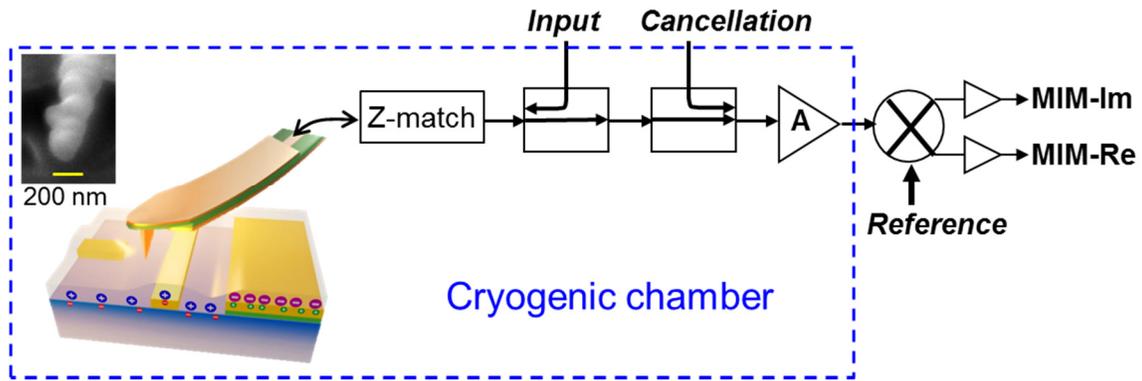

Fig. S1. Schematic of the cryogenic MIM system. An impedance (Z) match section, two directional couplers, and a cryogenic microwave amplifier are mounted in the Janis ST-500 vacuum chamber. The inset shows the FIB tip apex and the scale bar is 200 nm.

The cryogenic MIM setup is schematically shown in Fig. S1. A focused ion-beam (FIB) deposited Pt tip with a diameter of ~200 nm, as shown in the inset, was used in this experiment [S1]. During the experiment, the 1 GHz microwave signal is fed to the cantilever probe through a directional coupler and an impedance (Z) match section. The reflected signal is first combined with the cancellation signal through a second directional coupler before being amplified by a low-noise cryogenic amplifier. The output microwave signal is then demodulated by a room temperature mixer to form the MIM-Im and MIM-Re images.

**Section 2: Transport and MIM data on a different ion-gel-gated EDLT sample.**

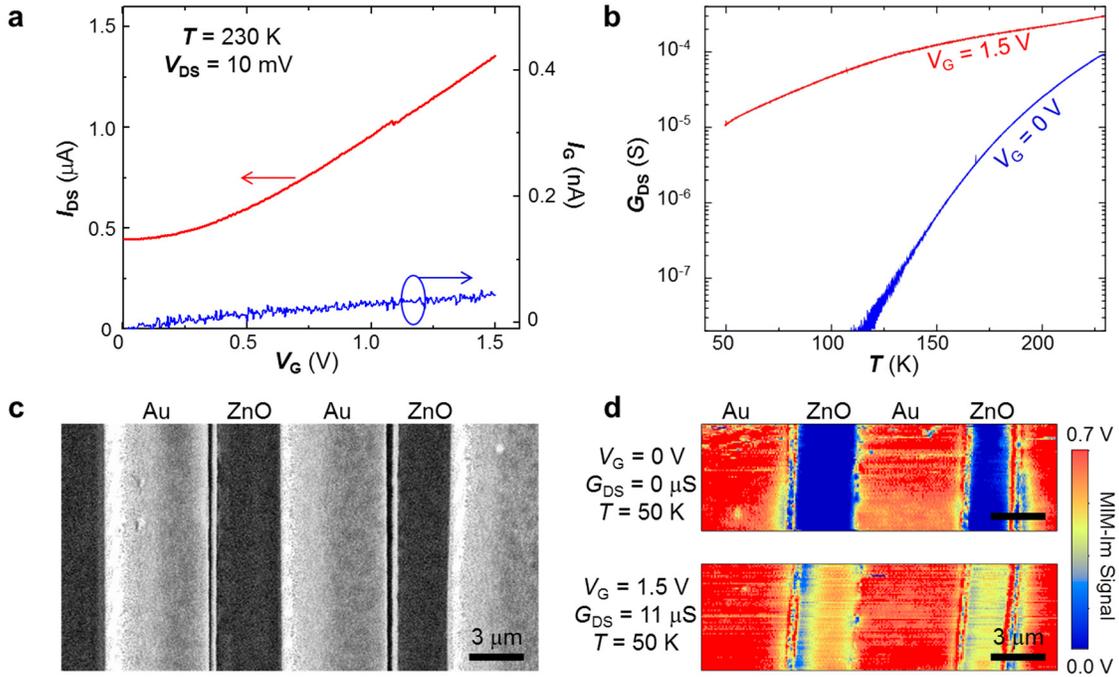

Fig. S2. (a) Transfer characteristics and leakage current of an EDLT device with parallel electrodes. (b) Temperature dependence of the source-drain conductance at the on ($V_G$ = 1.5 V) and off ($V_G$ = 0 V) states. (c) SEM and (d) MIM images of the device. The MIM data were taken at 50 K. All scale bars are 3 μm.

The T-shaped device is discussed in detail in the main text as the channel is mostly covered within one frame of MIM scans. On the other hand, the underlying physics does not rely on device configurations and we have also studied thin ion-gel-gated EDLTs with other contact geometry. Fig. S2(a) shows the transfer characteristics and leakage current of another EDLT device with parallel finger contacts [see Fig. S2(c)]. The device was turned on at $V_{th}$ ~ 0.3 V with a small gate leakage current $I_G$ < 0.1 nA. Note that $V_{th}$ and $I_G$ may vary from sample to sample due to the different ion-gel thicknesses. The metal-insulator transition was observed in the temperature dependence measurement Fig. S2(b). The corresponding MIM images acquired at 50 K in Fig. S2(d) show much higher signals at the ZnO channel in the on-state ($V_G$ = 1.5 V, $G_{DS}$ = 11 μS) than that of the off-state ($V_G$ = 0 V, $G_{DS}$ = 0 μS). The results agree well with that from the T-shaped device described in the main text.

**Section 3: Selected line profiles of the MIM images.**

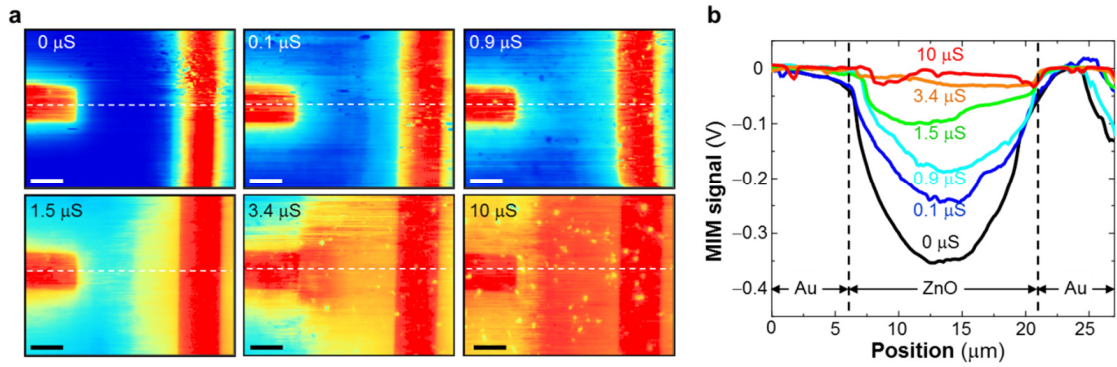

Fig. S3. (a) MIM images labeled by the two-terminal conductance $G_{DS}$ at 100 K, reproduced from Fig. 2(b). (b) Line cuts of the MIM images through the center of the source electrodes, shown as white dashed lines in (b). The regions of Au contacts and ZnO channel are indicated in the plot. All scale bars are 5 μm.

The MIM images labeled by the transport $G_{DS}$ at 100 K are reproduced here in Fig. S3(a), clearly showing the spatial evolution of metal-insulator transition as increasing channel conductance. We have also included a number of line scans through the center of the source electrodes in Fig. S3(b). Note that the MIM signal has a long tail when the ZnO is insulating, as the E-field decays slowly without screening from mobile charges. The effect is weaker in the crossover and conductive regions.

**Section 4: Full analysis of the MIM images.**

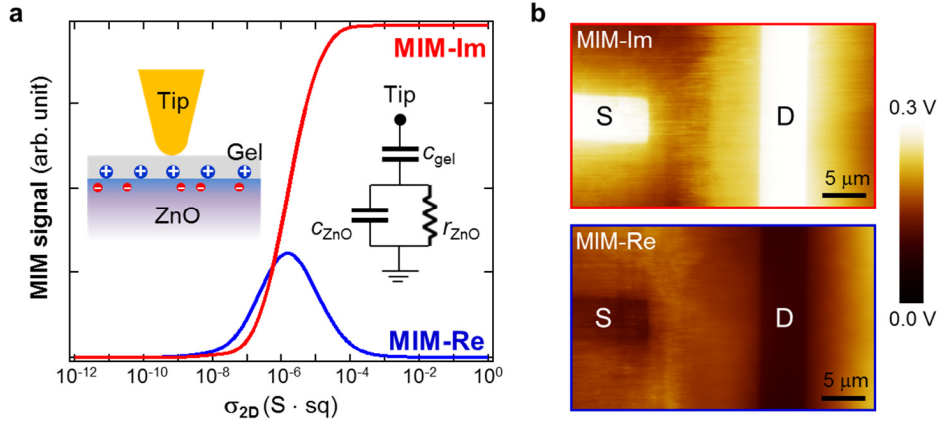

Fig. S4. (a) FEA simulation of the real (MIM-Re, blue) and imaginary (MIM-Im, red) components of the tip-sample admittance. The insets show the tip-sample configuration (left) and a lumped-element circuit model (right). (b) MIM-Im and MIM-Re images at $G_{DS}$ = 1.5 μS. Note the different false-color scale from the main text. Scale bars are 5 μm.

The tip-sample interaction can be modeled using the commercial FEA software COMSOL 4.4 [S1]. For the simulation, the diameter of the focused ion-beam (FIB) deposited Pt tip is assumed to be 200 nm [S1] and the thickness of the ionic gel is 50 nm. The dielectric constant (ε) of ZnO is 8.5 [S2]. While little is known on the dielectric constant at 1 GHz for the ionic gel, we have assumed here ε ~ 3, similar to typical polymers [S3]. The simulation result does not depend strongly on the dielectric constant. The thickness of the ZnO surface inversion layer is assumed to be $d$ = 5 nm and the 2D sheet conductance $\sigma_{2D} = \sigma_{3D} \cdot d$, where $\sigma_{3D}$ is the 3D conductivity.

The real (MIM-Re) and imaginary (MIM-Im) parts of the effective tip admittance, which are directly proportional to the MIM signals, are shown in Fig. S4(a). The results can be qualitatively understood by the lumped-element circuit in the inset [S1]. Note that this equivalent circuit seen by the tip is for the 1 GHz microwave excitation, which is totally different from the DC effective circuit in the inset of Fig. 1(c). When the gel-ZnO interface is insulating, the resistance at the ZnO surface $r_{ZnO}$ is very large and the tip is loaded by two geometric capacitors ($c_{gel}$ and $c_{ZnO}$) in series. When the gel-ZnO interface is highly conducting, the small $r_{ZnO}$ effectively shunts $c_{ZnO}$. In between these two limits, the MIM-Im signal increases monotonically as increasing $\sigma_{2D}$ and the MIM-Re signal reaches a peak around $\sigma_{2D}$ = 1 μS · sq. For simplicity, only the MIM-Im component is presented in the main text. A typical MIM-Re image [$G_{DS}$ = 1.5 µS in Fig. 2(b)] is shown in Fig. S4(b). Note that the metal electrodes have low MIM-Re signals due to their high conductivity.

In different cool-downs of the devices, particles with low MIM signals may or may not be observed on the surface. For instance, little evidence of surface particles is seen in the images in Fig. S4(b), while they are clearly visible in Figs. 4(b) and 4(e). The origin of these features is not entirely clear. It is possible that they are small ice particles due to water molecules dissolved in the ionic gel. Their locations could change if the device was thermally cycled to 300 K but would remain fixed if only cycled to 230 K, e.g. for the three images shown in Fig. 3(b) in the main text. Note that we always kept $V_G$ = 0 V if the system was warmed above 230 K. The transport results were very repeatable from cool-down to cool-down, suggesting that irreversible electrochemical effects are not important in this experiment.

**Section 5: Poisson-Schrodinger calculation at different surface band bending values.**

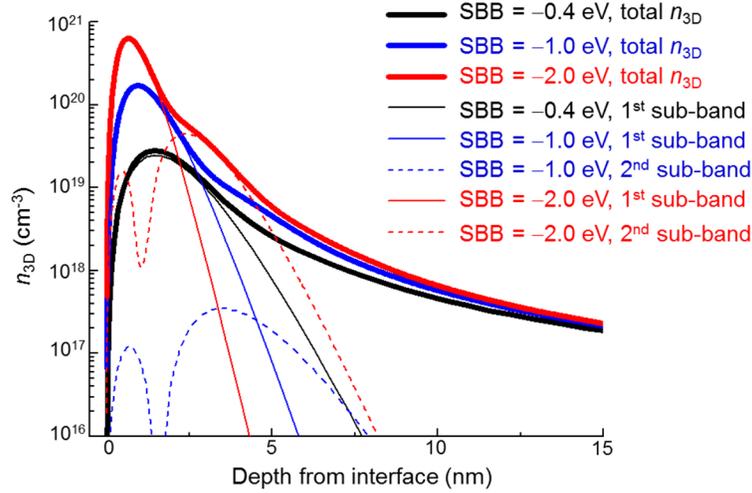

Fig. S5. 3D electron density ($n_{3D}$) as a function of the depth from the EDLT interface. The dashed lines show the contribution from the 1st and 2nd sub-bands. The temperature is 300 K in this simulation.

Fig. S5 shows the results of the self-consistent Poisson-Schrodinger simulation using a non-uniform-meshed method described in Ref. [S4]. The mesh is denser close to the interface to enhance accuracy and efficiency. In order to simulate the experiment in a range of temperatures, Fermi-Dirac distribution is also incorporated, so that thermally excited electrons in sub-bands with energy higher than the Fermi level are also taken into account. The effective mass is adapted from Ref. [S5] and the dielectric constant used is from Ref. [S2]. The solving range in depth is from 0 to 100 nm to ensure wide enough space for realistic solutions. We assume in the simulation a uniform electron doping at the level of $1 \times 10^{16}$ cm$^{-3}$ in the whole sample.

Based on the effective 2D density $n_{2D}$ as a function of surface band bending (SBB), we can perform self-consistent modeling to obtain the conductance distribution in Fig. 5(d). Here we assume a much larger gate area than the channel region such that $V_G$ drops mostly on the gel-ZnO interface. In addition, the mobility of ZnO (100 cm$^2$/V·s) [S6], is assumed to be independent of the electron density. The conductance distribution affects the potential drop inside the channel, which results in different SBB at each point. The calculation runs iteratively until the change of potential distribution is less than 0.01 V.

**References:**

[S1] Lai, K., Kundhikanjana, W., Kelly, M. A. & Shen, Z.-X. Modeling and characterization of a cantilever-based near-field scanning microwave impedance microscope. *Rev. Sci. Instrum.* **79**, 063703 (2008).

[S2] Yoshikawa, H. & Adachi, S. Optical Constants of ZnO. *Jpn. J. Appl. Phys.* **36**, 6237 (1997).

[S3] http://webhotel2.tut.fi/projects/caeds/tekstit/plastics/plastics_PMMA.pdf

[S4] Tan, I.-H., Snider, G. L., Chang, L. D. & Hu, E. L. A self-consistent solution of Schrödinger-Poisson equations using a nonuniform mesh. *J. Appl. Phys.* **68**, 4071 (1990).

[S5] Tsukazaki, A., Ohtomo, A., Kita, T., Ohno, Y., Ohno, H. & Kawasaki, M. Quantum Hall Effect in Polar Oxide Heterostructures. *Science* **315**, 1388 (2007).

[S6] Yuan, H., Shimotani, H., Tsukazaki, A., Ohtomo, A., Kawasaki, M. & Iwasa, Y. High-Density Carrier Accumulation in ZnO Field-Effect Transistors Gated by Electric Double Layers of Ionic Liquids. *Adv. Funct. Mater.* **19**, 1046–1053 (2009).